# Using a Large Language Model as Design Material for an Interactive Museum Installation

Maria Padilla Engstrøm

IT University of Copenhagen, maria.padilla.e@googlemail.com

Anders Sundnes Løvlie

IT University of Copenhagen, asun@itu.dk

We present a work in progress that explores using a Large Language Model (LLM) as a design material for an interactive museum installation. LLMs offer the possibility of creating chatbots that can facilitate dynamic and human-like conversation, engaging in a form of role play to bring historical persons "to life" for visitors. However, LLMs are prone to producing misinformation, which runs counter to museums' core mission to educate the public. We use Research-through-Design to explore some approaches to navigating this dilemma through rapid prototyping and evaluation and propose some directions for further research. We suggest that designers may shape interactions with the chatbot to emphasize personal narratives and role play rather than historical facts or to intentionally highlight the unreliability of the chatbot outputs to provoke critical reflection.

## 1 INTRODUCTION

There is much debate in Human-Computer Interaction (HCI) about how to use Machine Learning (ML) as a "design material" [10]. It has been suggested that machine learning is uniquely difficult to design with [1,4,13]. However, due to the great success of Large Language Models (LLMs) in recent years there is a growing interest in using LLMs for a broad variety of applications, including for the design of visitor experiences in museums [3,5,12,15,16,23–25,27], sometimes in the form of chatbots [17,21,26].

Chatbots have been used in museums for many years, using a variety of techniques. Chatbots may be used to expand the interactivity of a visit, offer a broader variety of information and interaction possibilities, and personalize the experience. However, a common challenge with the design of chatbots for museums is that "the designed conversations are not human-like, lacking meaningful interaction and engagement, and that limited knowledge is provided in response to user's questions" [21]. The use of chatbots based on LLMs may offer richer and more dynamic conversations, however, LLMs have a weakness: They may output information that is inaccurate, context-insensitive, or offensive, as they are prone to so-called "hallucination" [11,14,17]. This is a major obstacle to the use of LLMs in museums, as educating the public is a core tenet of museums, and disseminating inaccurate information would run counter to the museum's core mission.

In this paper, we report on a pilot study using Research-through-Design [7,28] to explore the design of a chatbot for The Worker's Museum in Copenhagen, which presents the history of the labour movement in Denmark. Through practical experiments using ChatGPT as a prototyping tool, we explore how a chatbot could be used to personify historical persons and enable the experience of having a (simulated) conversation with people from the past, stimulating curiosity and engagement while mitigating the risk that the chatbot might share inaccurate information. We identify a central dilemma in avoiding that measures taken to safeguard accuracy of information also might stifle the liveliness of the interactions, defeating the purpose of using a chatbot in the first place. Furthermore we suggest some directions designers might explore in order to deal with this dilemma.

## 2 RELATED WORK

Museums use chatbots in various ways across diverse platforms. Among the more common chatbots are those also known as infobots, which typically function as guides to help visitors answer questions about their museum visit or the museum's exhibitions and objects [20]. Chatbots have been created with a variety of technologies and used for a range of purposes, such as providing Q&A services [22], offering visitor information combined with personal narratives [6,20,21], and combining visitor guides with interactive quizzes and games [2]. However, several researchers have pointed out that many of the chatbots currently in use at museums are often limited in their knowledge and ability to engage in meaningful human dialogues, resulting in unengaging interactions [6,21].

With the recent advancements in large language models, new opportunities are emerging to design chatbots that offer the possibility of engaging in more flexible and human-like conversations with museum visitors. Scholars have explored using LLMs to enrich information exploration in museum websites [3], and for offering personalized tours [23,26,27]. Multimodal models have been used for purposes such as visual question answering [17,24], exploring artistic styles [12], and exploring artwork details [15]. However, LLMs are prone to "hallucination", which refers to the generation of false or inaccurate information [11,14]. Mountantonakis and colleagues [16] propose combining LLMs with Knowledge Graphs (KG) to mitigate the risk of LLM hallucination. Tonmoy and colleagues [19] survey the state of the art of mitigation techniques; however, avoiding hallucination in LLMs remains a tough challenge. A major European art museum recently deployed a chatbot to personify a historical artist, using Retrieval-Augmented Generation to safeguard against hallucinations; however, this caused the chatbot interactions to be inhibited to the degree that the chatbot could only answer a small range of questions, resulting in interactions which were described by one influential art critic as "profoundly meaningless" [9].

## 3 METHOD

We report on a pilot study using the methodology of Research-through-Design [7,28], in which insights are derived from the process of designing an interactive artifact. The study has been carried out with a minimum of resources and has therefore employed rapid prototyping with off-the-shelf technology (ChatGPT) to gain early insights into the potentials and challenges involved. Test users have been recruited from our networks using snowball sampling, allowing us to test the prototypes on potential users in our primary target audience as well as museum experts and collect input from a diverse group of users and stakeholders.

## 4 DESIGN

Our design project was carried out in collaboration with The Worker's Museum, whose core vision and strategy plan sets out an ambition "to create engaging encounters with history". Central to this ambition is creating visitor experiences that activate the visitors in novel ways, e.g., through interactive, multisensory, and embodied experiences. Through initial contextual research at the museum, we identified an interesting design challenge in a section of the museum presenting the history of striking workers from the late 18th century, represented by a group of physical mannequins dressed up in historical garments and carrying flags of the worker's movement. Our primary target audience was young museum visitors aged 13-15 years.

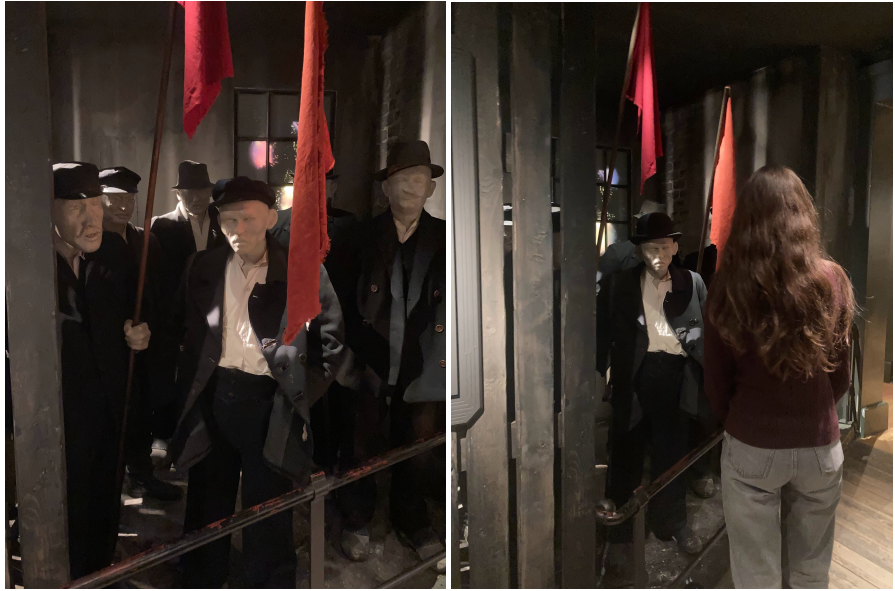

Figure 1: Left: A display of mannequins which represent workers during a strike in the late 19th century. Right: One of the test participants during a Wizard of Oz test.

Although the mannequins were intended to bring history to life for museum visitors by illustrating what a striking worker group might have looked like at the time, they appeared static and somewhat lifeless. We observed that most museum visitors passed by them quickly without stopping. We identified an interesting opportunity in using an interactive chatbot to bring these mannequins to life, allowing museum visitors to engage in dialogue with them. Being mindful of the danger of LLM "hallucinations", we set out to prototype and explore how a chatbot could be implemented to facilitate curiosity and engagement while respecting the museum's need for accuracy and veracity.

Our experiments proceeded through two main iterations, using ChatGPT as an easily accessible technology for rapid prototyping. In the first iteration, we wanted to explore whether we could use ChatGPT to create historical "personas" that would enact the role of the historical people represented by the mannequins - while avoiding inaccurate responses through restrictions given in an initial prompt. We created prompts for three personas: 'Erik', 'Jørgen', and 'Lars'. For each persona, the LLM was instructed to respond as a working-class man from the 18th century standing in the middle of a demonstration, giving answers based on historical background materials published by the museum which present the period considered as the start of the worker's movement in Denmark during the second half of the 19th century. The materials describe the typical working conditions of the workers and tell the chronological events leading up to the large strike represented by the mannequin display. The three personas differed from one another regarding their instructions for responding to questions that could not be answered based on the provided materials. 'Lars' was instructed to provide a standardized response whenever he could not find the necessary information in the material: "I don't know anything about that. Please try instead to ask me a question about the things I am interested in." 'Jørgen' was instructed to follow a different rhetorical strategy: Instead of responding directly to the question, the persona should redirect the conversation and try to offer an answer about a relevant topic covered by the historical material. Finally, 'Erik' received no instructions on how to handle inquiries not covered by the historical background materials.

To assess the feasibility of our approach, we carried out a series of user tests with a total of 10 participants. 8 of these were young people recruited through our networks, aged 12-28, 6 females and 2 males. The final 2 test users were employees in the museum's education team, whom we invited to capture the museum staff's assessment of the outputs from the three different personas. All the tests were carried out using the Wizard of Oz technique. The test participants were instructed to speak directly to the mannequins as they would to a living person. A researcher using a laptop computer typed the participants' questions into ChatGPT and used the "read aloud" function of ChatGPT to have the question read aloud to the participants.

The personas received 108 questions in total during the tests. When analyzing the transcripts of the conversations, we found zero cases where the personas offered factually incorrect information - which is encouraging, considering commonly cited concerns about LLM hallucinations. However, we found 10 cases altogether where the personas deviated from the instructions given in the initial prompts. In multiple test sessions, the personas stated that they were giving information that was based on the sources we had uploaded from the museum's website, even though the stated information did not exist in the sources. Thus, our test confirmed the previously stated concerns about the accuracy of information provided by ChatGPT.

Considering the participant's experience, it was striking that although the museum display presented the workers as a group rather than as particular individuals, the test participants quite consistently addressed the personas as individuals and asked questions about their opinions, their feelings, and their everyday lives. Most of the participants preferred talking to Erik - the one who did not have any restrictions on answering - as his responses were more engaging and expressive. Lars turned out to be unable to answer many questions, even mundane questions such as "What is your favorite food?" and "Do you have any pets?". Interestingly, while the two museum professionals before the interaction emphasized their concerns regarding the risk of disinformation, in their interactions with the personas it was clear that they much preferred interacting with Erik. Interacting with Lars quickly frustrated them as they repeatedly received the standardized response, similar to the younger participants. Thus, it seems that a central challenge would be to find a good balance between measures to mitigate against inaccurate information on the one hand and the need to facilitate lively conversations on the other.

Given the test participants' interest in asking personal questions, we carried out a second iteration of rapid prototyping to explore how to further deepen a more personal narrative, by creating two new personas: One, named 'Mads', was provided not just with factual historical materials, but also with a historical novel from the beginning of the 19th century, and was instructed to base his answers on the story of the protagonist of this novel. The second persona, named 'Jens', was instructed to answer questions based on a collection of factual articles describing the historical period. In order to focus on our primary target audience, we carried out a small test on three girls, all aged 14 years. As in the first iteration, all participants agreed that they preferred to hear about the personas' personal stories and experiences rather than the more general description of the period. However, they also found it interesting when factual descriptions of the period were incorporated as part of the personal story. In the below example are answers from Jens and Mads to the same question, asked by one of the participants: "What was your childhood like?"

> Jens: "My upbringing was marked by poverty and hard work. I grew up in a working-class family where we all had to help out from an early age to put food on the table. We lived close together in small, cramped apartments without much space or comfort. It was a struggle to get by, but we learned to stick together and do what we could to survive."

> Mads: "My upbringing was tough. I grew up in Sweden with my father, Lasse, after my mother died when I was little. We worked as day laborers on farms, where we toiled for little food and shelter. It was a poor life, but my father taught me the value of hard work and holding on to hope for a better future. That's why we traveled to Denmark to start a new life."

The participant indicated that she preferred Mads' answer, because it offered more personal details and made him seem more "human" and more interesting to talk to. However, in general the participants could not clearly identify how Mads and Jens answered questions differently. This seems to have been caused by 'Jens' sometimes deviating from the instructions and offering personal information not included in the background material (e.g., talking about his children and family life). Out of 38 questions, we found 5 responses where the personas deviated from their instructions.

## 5 DISCUSSION

Our study demonstrates some early insights into the challenges of using LLM as a design material for the historical museum context. While the prototypes did not provide factually incorrect information, all prototypes at times deviated from the instructions in their prompts. Meanwhile, the test participants' reactions demonstrate the dilemma between guarding against misinformation without stifling the possibility of having lively and engaging conversations with the chatbots. There is a need for more extensive experiments to explore these issues more deeply - using LLMs that have been customized for the purpose, perhaps in combination with other techniques. Based on our preliminary results, we outline below some directions that such experiments might pursue.

Designers may wish to deploy a chatbot in a way that guarantees that the outputs from the chatbot are historically accurate. This is very hard to achieve with an LLM, as the propensity for hallucinations appears to be a deeply entrenched property of such models. It is possible to provide better "guardrails" than in our simple pilot study, for instance by using Retrieval-Augmented Generation and similar techniques [19]. However, such approaches run the risk of too strongly inhibiting the conversations with the chatbots, which might defeat the purpose of creating a chatbot - if that purpose was to increase interactivity and engagement.

An alternative strategy might be to embrace the tendency demonstrated by test users in our pilot study to focus on personal narratives and design chatbots for conversations that emphasize the personal and emotional, focusing more on facilitating empathy with people of the past than communicating historical facts. Such a design strategy could mitigate the risk of historical inaccuracy by steering conversations away from facts and instead highlighting the personal and emotional - drawing on the role-playing characteristic of LLMs [18]. Furthermore, such designs might communicate to users that they are presenting a narrative with elements of fiction and imagination, thus cautioning users away from interpreting the chatbot's statements as reliable statements about historical facts. Such a strategy was pursued by some of our students when designing an installation for The National Museum of Denmark, in which a chatbot was used to impersonate a Viking sorceress who would tell the fortunes of museum visitors. In such a design, the factual veracity of the chatbot's statements is de-emphasized, inviting visitors to consider the chatbot outputs not as reliable statements of fact but rather as a fictional role-playing of magical rituals carried out in the time of the Vikings. Such a strategy could be applied to the The Worker's Museum for instance by creating chatbots that represent different stakeholders in the conflict between workers and factory owners, and have these chatbots present their perspective on the conflict. In such a scenario it would not be crucial that the chatbot is always able to produce fully reliable facts - we might not trust in everything the participants in the conflict say - but rather, what would be important would be the chatbot's ability to play its role in a believable way and stick to the role throughout a conversation.

Finally, designers might choose to actively play on the unreliability of LLMs to design conversations that provoke critical reflection by highlighting the fact that the LLMs' outputs can not be trusted. Such designs might use implicit techniques similar to the way a narrator in fiction and film may be revealed as unreliable; or a design might explicitly present itself as questionable or unreliable, inviting users to reflect on the complexity of historical interpretation, for instance by presenting multiple interpretations of the same materials. Such designs might mimic some of the techniques for ambiguity suggested by [8]. Applying this strategy to the The Worker's Museum we might for instance create a scenario where different factions within the labor union were arguing their case. The visitor could talk to chatbots representing social democrats, communists, anarchists, conservatives and other political factions, while being invited to think critically about the claims of these political faction leaders. Or visitors might be invited to chat with a factory owner who defends the working conditions in the factory, which were clearly unacceptable to a contemporary citizen; a chatbot that defends something indefensible for a contemporary visitor would invite a critical response and the critical exchange would not require the chatbot to be always truthful. Furthermore, such exchanges with an unreliable chatbot might also be used to highlight uncertainty in museological interpretation. E.g. a museum might present some old objects for which there is no clear scientific consensus about their interpretation - a quite common situation in museology - and invite visitors to interact with chatbots that represent different interpretations regarding the objects, while being made aware that no one knows with certainty which interpretation is correct.

## 6 CONCLUSION

Evaluation of our prototypes has demonstrated that interacting with an LLM-based chatbot can be attractive and engaging for museum visitors, who tend to ask questions about personal stories rather than more general historical information. However, we see a core dilemma for designers in balancing the need to safeguard against LLM hallucinations against the need to facilitate lively and engaging interactions for users. Since safeguarding outputs from LLMs is very difficult, we propose that designers focus their efforts on setting the expectations of users and the premises for the interaction in ways that either steer the conversation away from historical facts, focusing instead on personal narratives, or in ways that make it clear to users that the outputs cannot always be trusted and invite critical reflection. Such design approaches have not been explored in the research literature discussing the use of LLMs in museums; however, given the great challenge of safeguarding LLM outputs, there is both a need and an opportunity to design the interactions in a way that allows visitors to have engaging experiences while being aware that the LLM may not always speak authoritatively or truthfully about historical facts.

## ACKNOWLEDGMENTS

We thank The Worker's Museum for facilitating the collaboration described in this paper. Further thanks to Christian Sivertsen for helpful comments on a draft version of this paper.